\documentclass[numberedheadings]{aip-cp}

\usepackage[numbers]{natbib}
\usepackage{rotating}
\usepackage{graphicx}
\begin{document}

\title{The Single Mirror Small Sized Telescope For The Cherenkov Telescope Array}

\author[addr1]{M.~Heller}
\author[addr1]{E.jr~Schioppa}
\author[addr1]{A.~Porcelli}
\author[addr1]{I.~Troyano Pujadas}
\author[addr5]{K.~Zi{\c e}tara}
\author[addr1]{D.~della Volpe}
\author[addr1]{T.~Montaruli}
\author[addr1]{F.~Cadoux}
\author[addr1]{Y.~Favre}
\author[addr1,addr13]{J.~A.~Aguilar}
\author[addr1]{A.~Christov}
\author[addr2]{E.~Prandini}
\author[addr9]{P.~Rajda}
\author[addr1]{M.~Rameez}
\author[addr9]{W.~Bilnik}
\author[addr3]{J.~B\l{}ocki}
\author[addr12]{L.~Bogacz}
\author[addr9]{J. Borkowski}
\author[addr4]{T.~Bulik}
\author[addr7]{A.~Frankowski}
\author[addr4]{M.~Grudzi{\'n}ska}
\author[addr5]{B.~Id{\'z}kowski}
\author[addr5]{M.~Jamrozy}
\author[addr7]{M.~Janiak}
\author[addr9]{J.~Kasperek}
\author[addr9]{K.~Lalik}
\author[addr2]{E.~Lyard}
\author[addr3]{E.~Mach}
\author[addr10]{D.~Mandat}
\author[addr3,addr5]{A.~Marsza{\l}ek}
\author[addr1]{L.~D.~Medina~Miranda}
\author[addr3]{J.~Micha{\l}owski}
\author[addr7]{R.~Moderski}
\author[addr2]{A.~Neronov}
\author[addr3]{J.~Niemiec}
\author[addr5]{M.~Ostrowski}
\author[addr6]{P.~Pa{\'s}ko}
\author[addr10]{M.~Pech}
\author[addr10]{P.~Schovanek}
\author[addr6]{K.~Seweryn}
\author[addr2,addr8]{V.~Sliusar}
\author[addr3]{K.~Skowron}
\author[addr5]{{\L}.~Stawarz}
\author[addr3,addr5]{M.~Stodulska}
\author[addr3]{M.~Stodulski}
\author[addr2]{R.~Walter}
\author[addr9]{M.~Wi{\c e}cek}
\author[addr5]{A.~Zagda\'{n}ski}
\author[]{the CTA consortium}

\affil[addr1]{DPNC - Universit\'e de Gen\`eve, 24 Quai Ernest Ansermet, Gen\`eve, Switzerland}
\affil[addr2]{D\'epartment of Astronomy - Universit\'e de Gen\'eve, 16 Chemin de Ecogia, Gen\`eve, Switzerland}
\affil[addr3]{Instytut Fizyki J\c{a}drowej im. H. Niewodnicza\'nskiego Polskiej Akademii Nauk, ul. Radzikowskiego 152, 31-342 Krak\'ow, Poland}
\affil[addr4]{Astronomical Observatory, University of Warsaw, al. Ujazdowskie 4, 00-478 Warsaw, Poland}
\affil[addr5]{Astronomical Observatory, Jagellonian University, ul. Orla 171, 30-244 Krak\'ow, Poland}
\affil[addr6]{Centrum Bada\'n Kosmicznych Polskiej Akademii Nauk, Warsaw, Poland}
\affil[addr7]{Nicolaus Copernicus Astronomical Center, Polish Academy of Science, Warsaw, Poland}
\affil[addr8]{Astronomical Observatory, Taras Shevchenko National University of Kyiv, Observatorna str., 3, Kyiv, Ukraine}
\affil[addr9]{AGH University of Science and Technology, al.Mickiewicza 30, Krak\'ow, Poland}
\affil[addr10]{Institute of Physics of the Czech Academy of Sciences, 17. listopadu 50, Olomouc \& Na Slovance 2, Prague, Czech Republic}
\affil[addr11]{Vrije Universiteit Brussels, Pleinlaan 2 1050 Brussels, Belgium}
\affil[addr12]{Department of Information Technologies, Jagiellonian University, ul.\ prof.\ Stanis{\l}awa {\L}ojasiewicza 11, 30-348 Krak\'ow, Poland}
\affil[addr13]{Universit\'e Libre Bruxelles, Facult\'e des Sciences, Avenue Franklin Roosevelt 50, 1050 Brussels, Belgium}

\corresp[cor1]{Corresponding author: matthieu.heller@unige.ch}

\maketitle

\begin{abstract}
The Small Size Telescope with Single Mirror (SST-1M) is one of the proposed types of Small Size Telescopes (SST) for the Cherenkov Telescope Array (CTA). About 70 SST telescopes will be part the CTA southern array which will also include Medium Sized Telescopes (MST) in its threshold configuration. Optimized for the detection of gamma rays in the energy range from 5~TeV to 300~TeV,  the SST-1M uses a Davies-Cotton optics with a 4~m dish diameter with a field of view of 9$^{\circ}$. The Cherenkov light resulting from the interaction of the gamma-rays in the atmosphere is focused onto a 88~cm side-to-side hexagonal photo-detection plane. The latter is composed of 1296 hollow light guides coupled to large area hexagonal silicon photomultipliers (SiPM). The SiPM readout is fully digital readout as for the trigger system. The compact and lightweight design of the SST-1M camera offers very high performance ideal for gamma-ray observation requirement.
In this contribution, the concept, design, performance and status of the first telescope prototype are presented.
\end{abstract}

\section{Introduction}
\label{sec:INTRO}
	\begin{figure}
		 \includegraphics[width=0.4\textwidth]{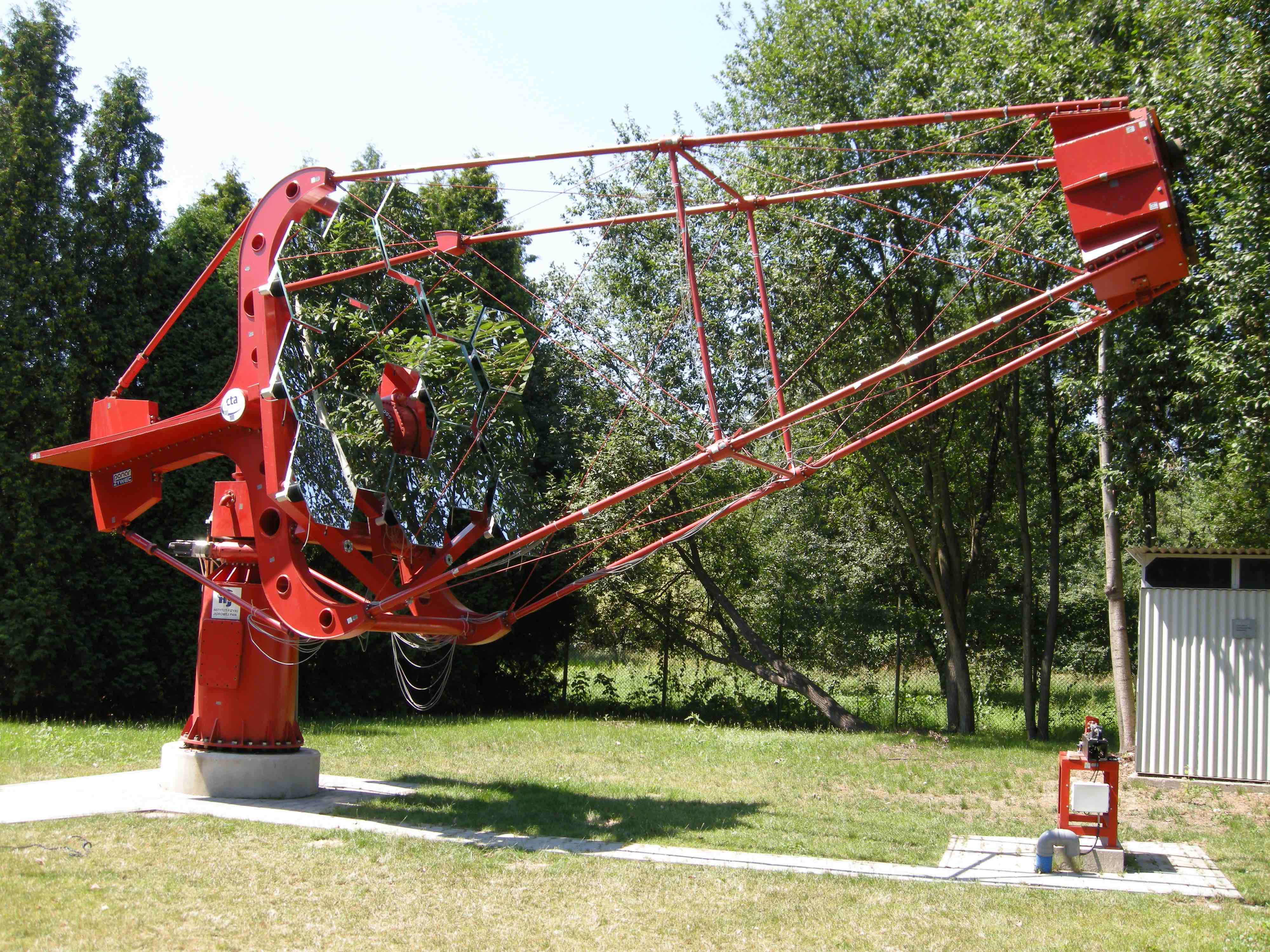}
		\caption{A picture of the SST-1M telescope structure installed at IFJ-PAN in Krak\'ow}
		\label{fig:TEL_STRUCT}	
	\end{figure}
The Cherenkov Telescope Array (CTA) \cite{CTAconcept} is a consortium of over 1300 members from more than 200 institutes from 32 countries proposing to build two arrays of more than 100 Imaging Air Cherenkov Telescopes (IACTs) in total to be located in two different sites in the south and north hemispheres. CTA will be the first open access observatory for gamma rays. The southern sub-array will cover the energy range between 20~GeV and 300~TeV. In order to reach such a wide energy spectrum, the array will be composed of telescopes of two different sizes, optimized for covering a specific sub-region of the energy spectrum: Medium Sized Telescopes (MST, with 12~m mirror diameter, 100~GeV-10~TeV) and Small Sized Telescopes (SST, with 4~m mirror diameter, 5~TeV-300~TeV).

Three different designs have been proposed for the SST to be installed exclusively in the southern sub-array. : two dual mirrors Schwarzschild-Couder telescopes \cite{ASTRI, GCT} and a single mirror Davies-Cotton telescope, the SST-1M, for a total of about 70 telescopes. The SST-1M project, presented here, is currently in its prototyping phase. The design of its various components has been carried out in order to optimize performance, costs and ease of installation in view of a possible production of up to 20 units for the final SST sub-array. A special care has been taken to adopt industrial standards for the production of each sub-elements of the telescope. The structure of the SST-1M prototype is shown in figure \ref{fig:TEL_STRUCT} and will be described in section \ref{sec:TelStruct}. The telescope optical system is described in section \ref{sec:optics}. The innovative camera based on silicon photomultiplier (SiPM) sensors will be described in section \ref{sec:camera}. Its completion at the University of Geneva is foreseen for the summer 2016 and its subsequent installation on the structure in Krak\'ow early fall 2016.

\section{The SST-1M telescope}
\label{sec:SST-1M_CONCEPT}
\subsection{The telescope structure and drive system}
\label{sec:TelStruct}
Figure \ref{fig:TEL_STRUCT} shows a picture of the telescope structure. Its design has been driven by the CTA design concept requirements: a point spread function for SSTs $\leq$ 0.25$^{\circ}$ at 4$^{\circ}$ off-axis; the telescope must focus light over 80\% of the required camera field of view diameter with an rms optical time spread of less than 1.5~ns; a field of view $\geq$ 8$^{\circ}$. These requirements determine the overall geometry of the telescope \cite{CONES}, including the camera pixels desbribed in section \ref{sec:camera}. Using a Davies-Cotton design, a 4~m reflector diameter with a focal ratio of 1.4 translates into a focal length of 5.6~m and satisfies the requirements.

The mechanics is mostly made of steel, and the majority of the components are built using off-the-shelf industrial products. The total weight of the telescope, including the camera, is 8.6 metric tons, making it relatively compact and lightweight. At the same time, the design confers the necessary stiffness for coping with the environmental and geological conditions (e.g. altitude, seismics, wind, etc) that can be expected on site. The main components of the structure are discribed in \cite{ICRC_struct}.

The telescope moves around the elevation (from -14$^{\circ}$ to 91$^{\circ}$) and the azimuth (from -270$^{\circ}$ to 270$^{\circ}$) axes by means of slew drives driven by two servo motors each, acting on worm gears. 
The drive system can provide a pointing accuracy within 7~arcsec and a tracking accuracy below 5~arcmin. The maximum angular velocities are 6$^{\circ}$/s in azimuth and 2$^{\circ}$/s in elevation. In fast slewing, the telescope can be pointed to any direction within one minute while the CTA requirement is 90~s.

\subsection{The telescope optical system}
\label{sec:optics}
The SST-1M reflector \cite{ICRC_mirrors} is composed of 18 hexagonal mirror facets of 78~cm flat-to-flat size. The facets are arranged into two concentric rings to form a spherical mirror, for a total area of 9.42~m$^{2}$. After including the shadowing from the telescope mast and the camera, and accounting for the mirror average reflectance (0.87$\pm$0.01 for wavelengths between 300~nm and 600~nm), the effective collection area reduces to 6.47~m$^{2}$.
\subsubsection{Mirrors}
\label{sec:Mirrors}
The spherical mirror facets have a radius of curvature of 11.2~m and must be produced in order to comply with the requirements in terms of reflectivity and mechanical and thermal stability. The prototype telescope is serving as a test bench to evaluate different mirror solutions. All mirrors underwent mechanical and optical tests demonstrating that the prototypes meet the CTA requirements \cite{SPIE_Enrico}.
\subsubsection{Alignment}
\label{sec:Alignment}
\begin{figure}
		 \includegraphics[width=0.5\textwidth]{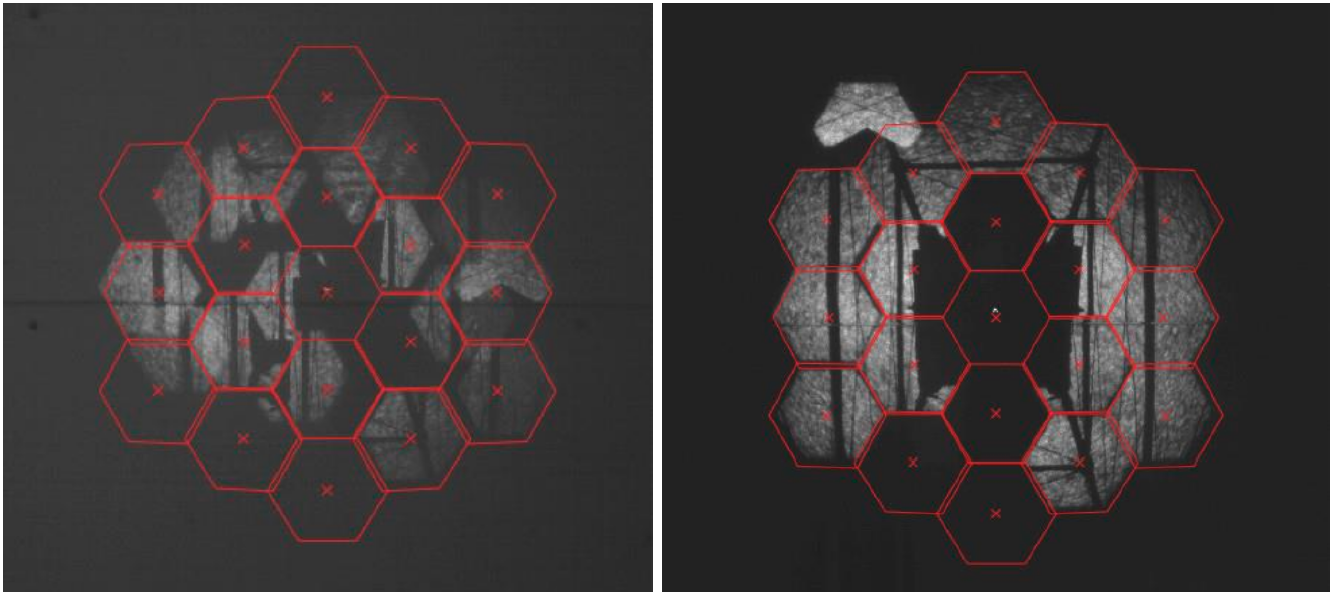}
		\caption{Manual test of the mirror alignment: picture of the reflector before alignment (left) and after alignment (right), except for the top mirror in group 1. Three mirrors where missing at the time of the test (visible in the bottom of the picture on the right).}
		\label{fig:MirrorAlignment}	
	\end{figure}

The alignment system uses an automatic optical feedback to fine-tune the relative positions of each facets in order to produce the required spherical surface focusing the Cherenkov light onto the focal plane. Each facet is installed on the dish via a fixing set hosting the actuators of the alignment system. The Bokeh alignment strategy \cite{BOKEH} used by the FACT collaboration has been used to perform the first manual alignment of the mirros facets. The result is shown in figure \ref{fig:MirrorAlignment}  where telescope was pointed to an external light source located around 45~m away. For the future, the NAMOD alignment strategy is concidered \cite{NAMOD}. In order to perform these techniques, a CCD camera installed in the center of the dish analyses the image of distant sources reflected on a surface installed on the camera lid. 
\subsection{The camera}
\label{sec:camera}
The SST-1M camera \cite{CameraPaper} uses a fully digital readout and trigger system (called DigiCam) that is physically separated from the photo-detection plane (optical system and its front-end electronics). Following the encouraging results from the FACT \cite{FACT, FACT2} in the use of silicon photomultipliers (SiPMs) for gamma-ray astronomy, the same technological solution was adopted. Indeed, SiPMs can be operated at relatively low bias voltage compared to standard PMTs and suffer from negligible ageing effects, so that they can be operated under high moon conditions without risks of degradation. As a consequence, the duty cycle increases enhancing the overall sensitivity of the full array in the high energy range. 
The camera dimensions in the focal plane (97~cm flat-to-flat for the full structure, 88~cm flat-to-flat for the photo-detection plane) are dictated by the telescope geometry, together with the CTA requirement of having a field of view above 8$^{\circ}$ for SSTs \cite{CONES}. The hexagonal-shaped enclosure (73.5~cm thick) hosts the photo-detection plane (PDP), DigiCam, the power supplies, the housekeeping  and the cooling system in a structure that complies at least with the IP65 standard, and which weighs less than 200~kg. 
The camera is sealed by a 3.3~mm thin Borofloat entrance window coated with an anti-reflective layer and a dichroic filter to cut-off light above 540~nm of wavelength, thus enhancing the signal-to-noise ratio of the camera. 
A water cooling system has been developed to extract the $\sim$2~kW of the camera (see Refs. \cite{SPIE_Enrico,CameraPaper}).
\subsection{The Photo-Detection Plane (PDP)}
\label{sec:PDP}
\subsubsection{The Pixels}
\label{sec:pix}
\begin{figure}
		 \includegraphics[width=0.25\textwidth]{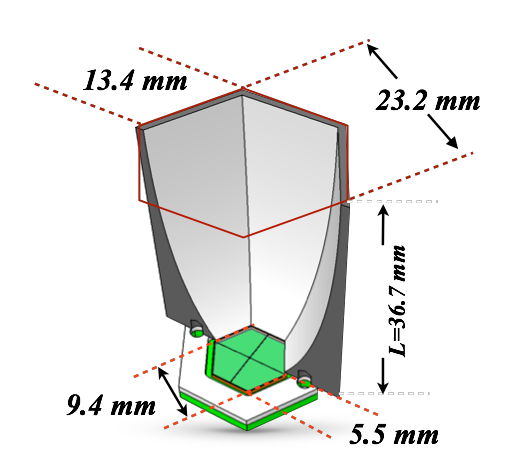}\hspace{0.3cm}
		 		 \includegraphics[width=0.25\textwidth]{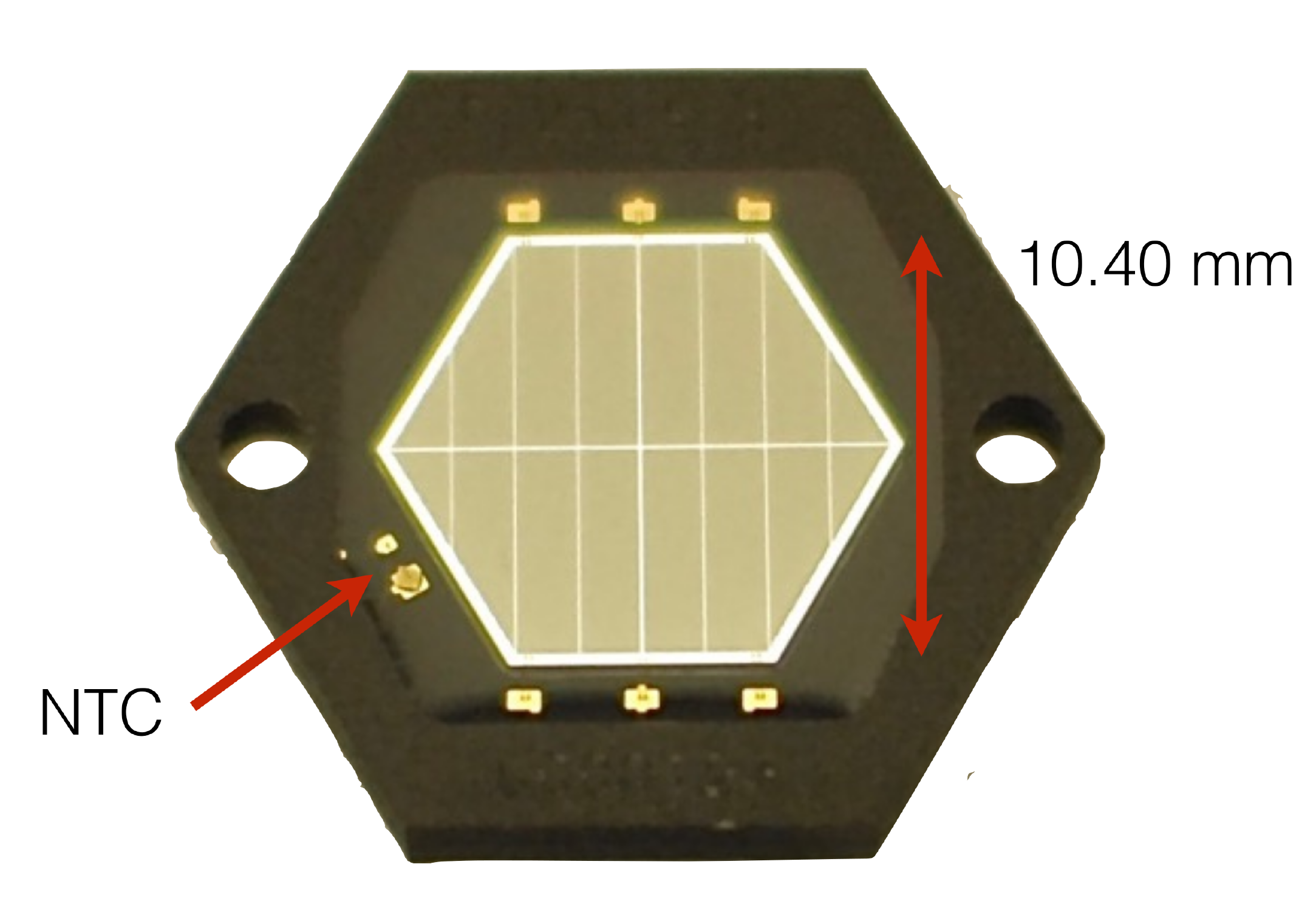}\hspace{0.3cm}
		 \includegraphics[width=0.25\textwidth]{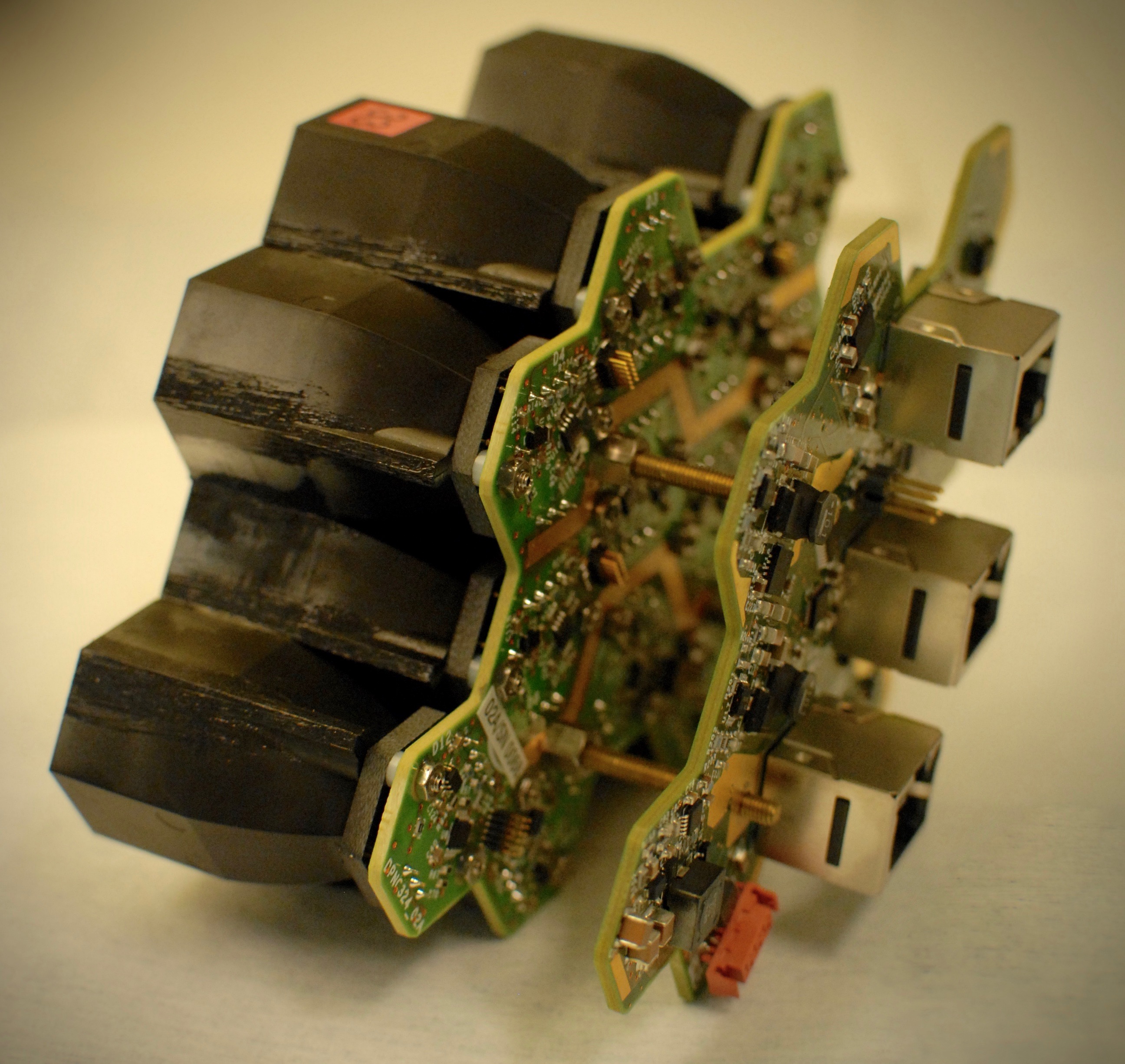}
		\caption{Left: Design drawing of the SST-1M pixel. Center: Detailed picture of hexagonal SiPM. Right: PDP Module made out of 12 pixels. The thermal foam inserted between the two PCBs and between the slow control board and the backplate are not visible in this picture}
		\label{fig:PDPdetail}	
	\end{figure}

The light reaching the entrance of the PDP  is focused onto the SiPM sensors by hollow light concentrators. The geometry of the concentrators has been optimized to reach the required 24$^{\circ}$ cut-off angle and 23.2~mm flat-to-flat pixel size on an hexagonal surface (see Ref. \cite{CONES}). The fabrication process (substrate and coating) has been optimized to enhance the reflectivity of the inner surface in the Cherenkov wavelength range. A detailed description of the production of the light funnels is given in Ref. \cite{CameraPaper}. Figure \ref{fig:PDPdetail} shows a drawing of a pixel unit (light concentrator plus sensor), highlighting the dimensions. The geometry of the light concentrator fixes the size of the underlying hexagonal SiPM to 9.4 mm side-to-side. 
Since they are not available in the market, custom designed sensors of this size and shape (93.56~mm$^{2}$) have been designed together with Hamamatsu in low cross-talk technology (see figure \ref{fig:PDPdetail}). Each sensor is provided with a embedded NTC temperature probe. The sensors have been extensively characterized and details can be found in Refs. \cite{CameraPaper,MPPC,ElectronicsPaper}.
As visible in figure \ref{fig:PDPdetail}, the 1296 pixels in the PDP are arranged into modules of 12 units each, for a total of 108 modules.
	\begin{figure}
		 \includegraphics[width=0.3\textwidth]{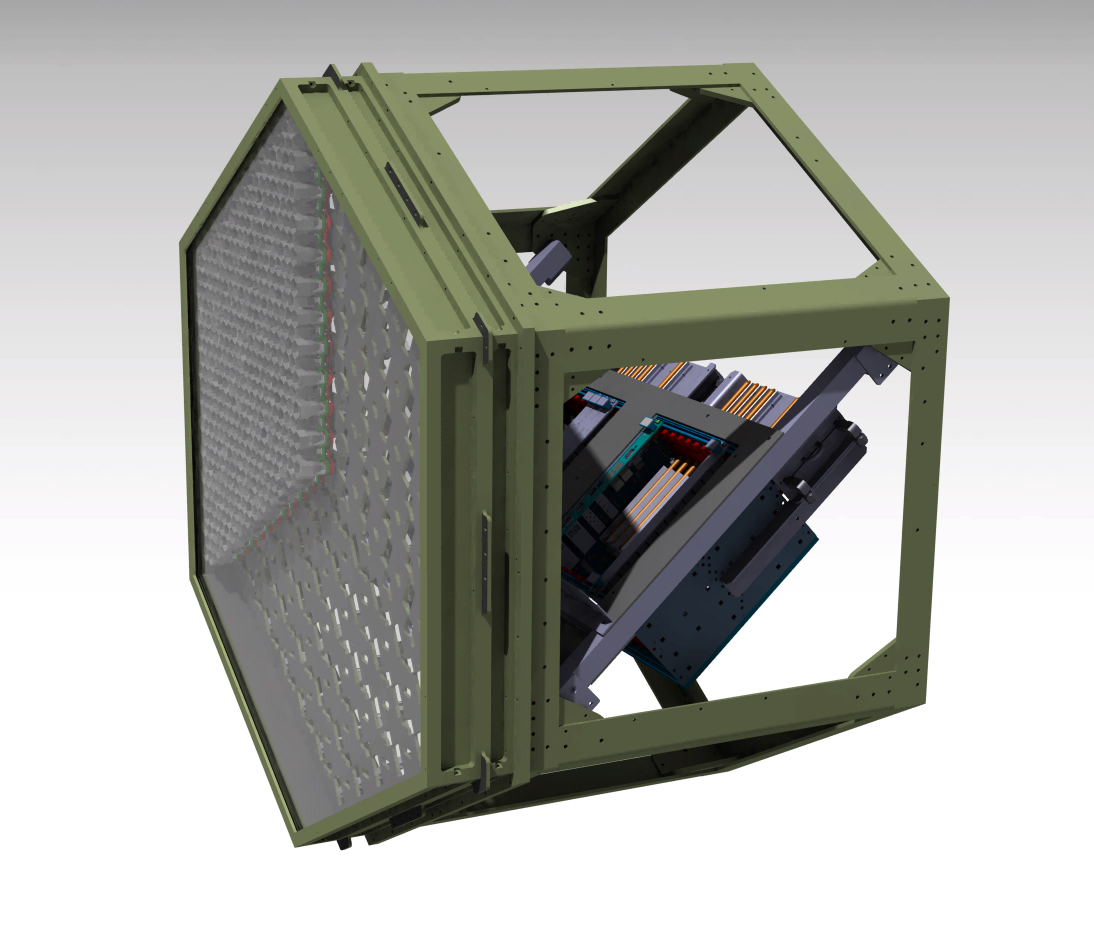}\hspace{1cm}
		 \includegraphics[width=0.275\textwidth]{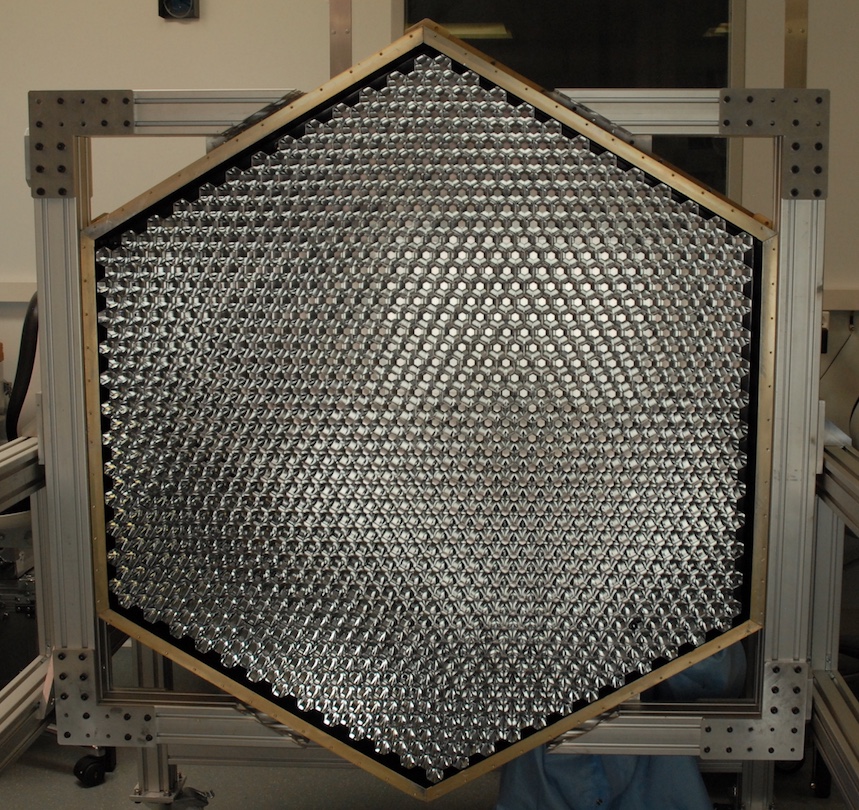}
		\caption{Left: 3D rendering of the camera. Right: Picture of the assembled PDP. At this stage, a Plexiglas window is used in order to protect the pixels and save the real window from scratches that could occur during the commissioning phase.}
		\label{fig:PDPass}	
	\end{figure}
\subsubsection{The front-end electronics}
\label{sec:FEE}
The camera front-end electronics consists of two main parts: the preamplifier board and the slow control board visible in figure \ref{fig:PDPdetail}. The 12 sensors of each module are soldered onto the preamplifier board, which hosts the 12 discrete preamplification stages. The unique size of the sensor made the design of the preamplifier stage rather challenging as detailed in Ref. \cite{ElectronicsPaper}. The sensor is DC coupled to the trans-impedance preamplifier topology. Despite the known disadvantages in term of dynamic range of the latter, it also offers the advantage of being able to monitor the night sky background continuously by measuring the baseline variations \cite{SPIE_Matthieu}. 
The slow control board routes the analog signals to the digital electronics, allow the monitoring through CAN bus protocol of the pixels' temperature and operational voltage and continuously adjust the bias voltage of the sensors according to the temperature measured by the embedded NTC probe.  This feature releases the constraint on the cooling system and ensure a stable operation point. Details and validation measurement have been shown in Refs. \cite{CameraPaper, ElectronicsPaper}.
\subsection{The Digital readout electronics DigiCam}
\label{sec:DigiCam}
\begin{figure}
		 \includegraphics[width=0.3\textwidth]{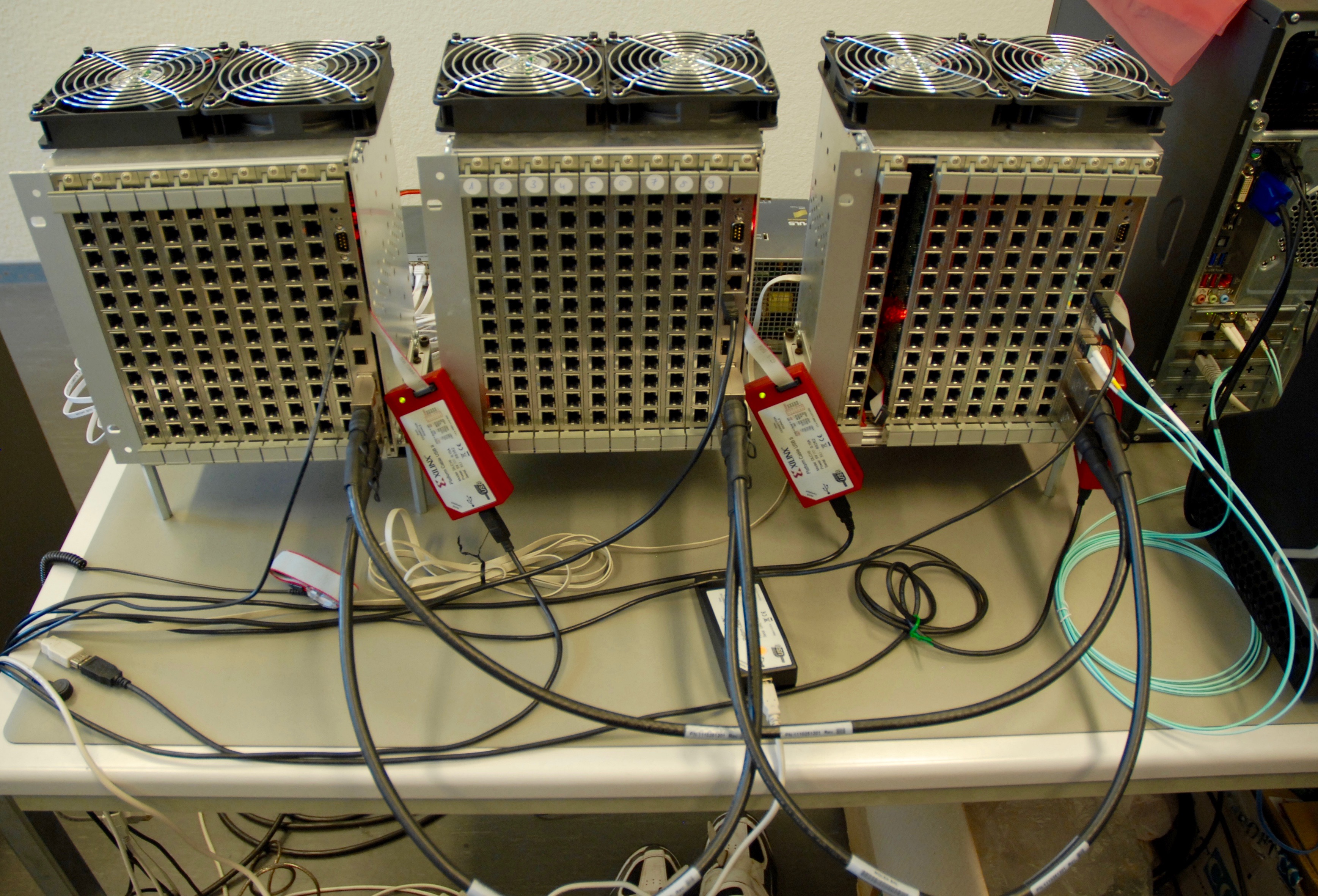}\hspace{1cm}
		 \includegraphics[width=0.27\textwidth]{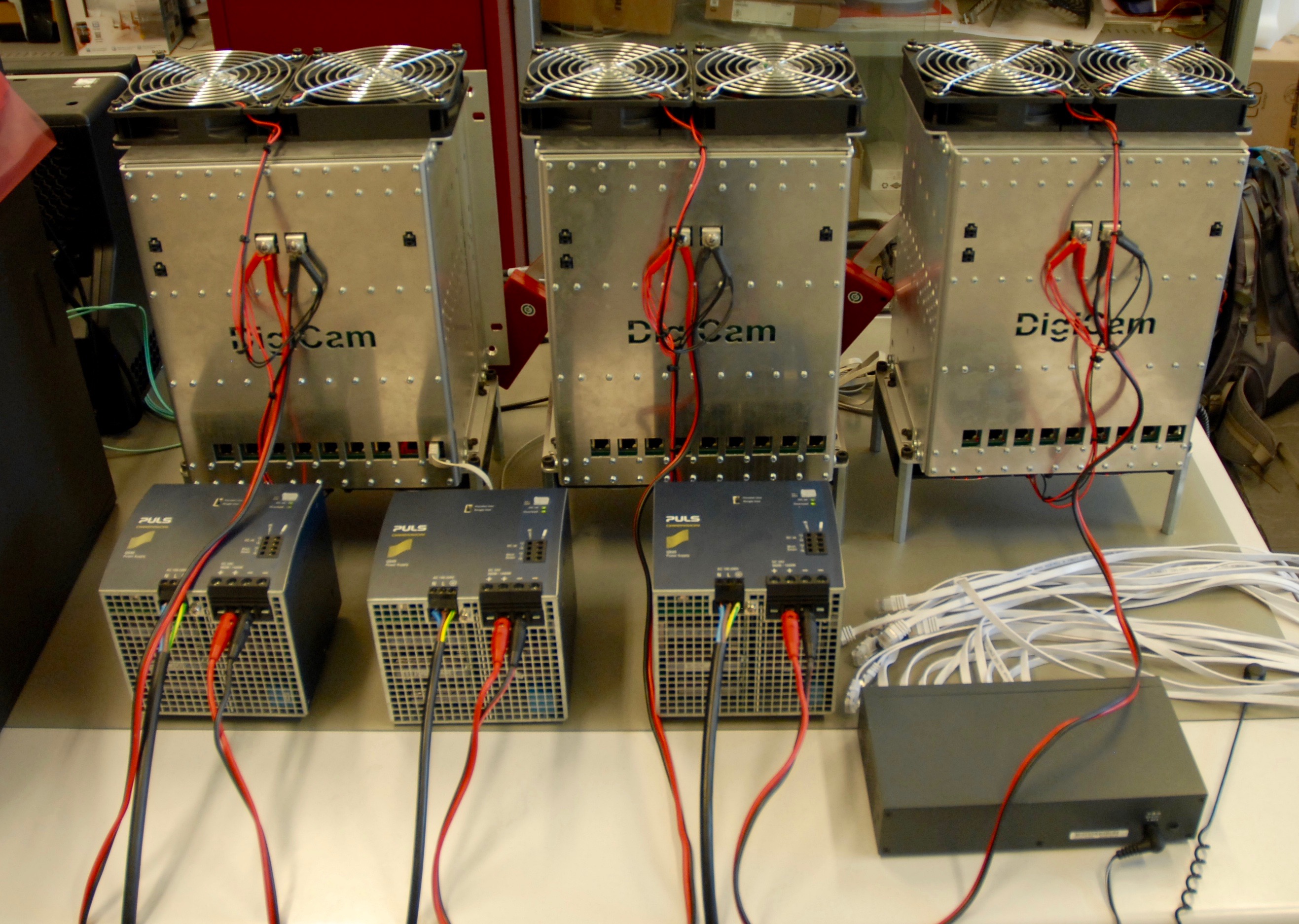}
		\caption{Picture of the 3 crates composing the DigiCam. Each crate readouts one sector of the camera and exchange their information through infiniband cables in order to be confronted to complex trigger topologies}
		\label{fig:DigiCam}	
	\end{figure}

The DigiCam hardware consists of three microcrates and their power supplies. The photo-detection plane is, in fact, divided into three logical sectors, each connected to one microcrate (see figure \ref{fig:DigiCam}). Each microcrate contains 9 fast digitizer boards (FADC) and one trigger board. Both components make use of the latest generation FPGA to reach high data throughput and to implement a highly flexible readout and trigger logic. Each FADC boards digitizes the signals from 4 modules (48 pixels) at a sampling rate of 250~MHz (4~ns sample size) on 12 bit digitizers. The digitized data are temporarily stored on ring buffers, while a low resolution copy, obtained by summing together the signals from groups of 3 adjacent pixels (patches) in 8-bit words, is sent to the trigger board of the microcrate. Here a highly parallelizable trigger algorithm is applied, based on the recognition of circular, elliptical and ring patterns (these latter typical of muon events) on patches that exceed a dedicated threshold. If an event is selected, the full resolution data are sent to the camera server via a 10 GB ethernet link. More details can be found in Refs. \cite{CameraPaper, ICRC_DigiCam}.

\section{The SST-1M performance}
\label{sec:perf}
\begin{figure}
	\includegraphics[width=0.45\textwidth]{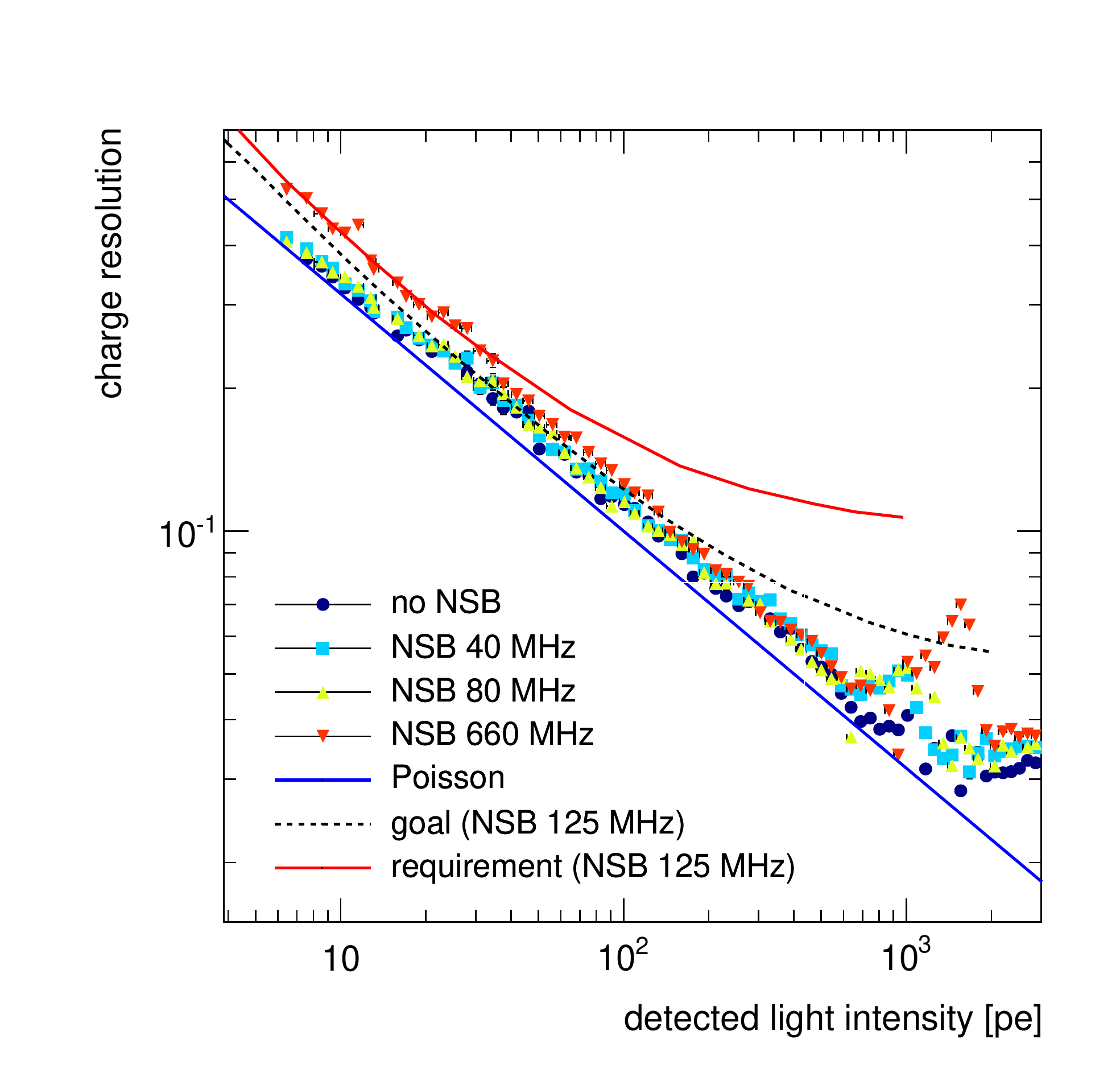}\hspace{0.5cm}
	\includegraphics[width=0.45\textwidth]{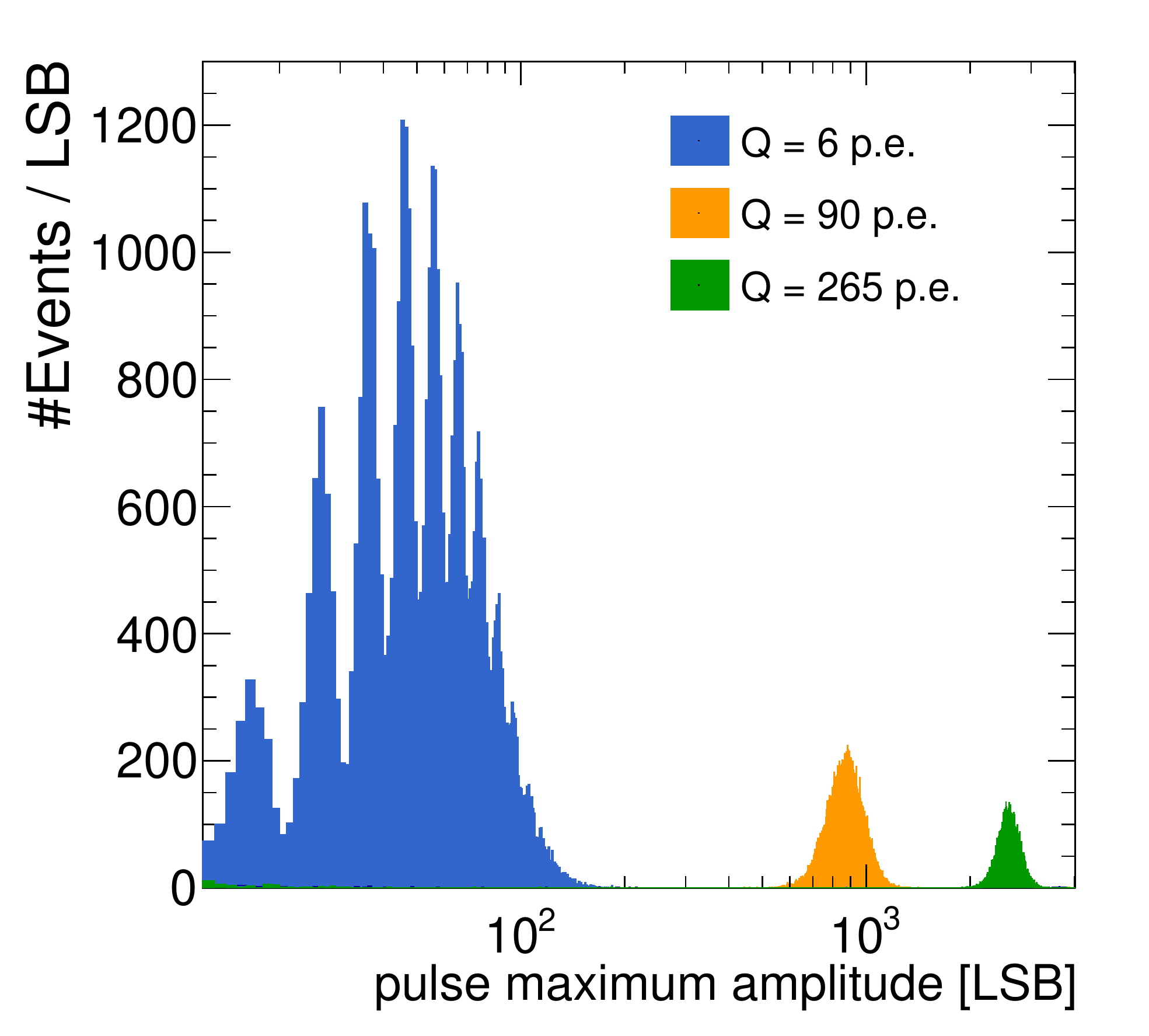}
		\caption{Left: The charge resolution measured in the dark (blue circles) and at different emulated night sky background levels: 40~MHz/pixel (cyan squares), corresponding to dark nights, 80~MHz/pixel (light green up-pointing triangles) and 660~MHz/pixel (red down-pointing triangles), corresponding to half moon nights with the moon at 4$^{\circ}$ off-axis. The data points are compared to the Poisson limit (blue continuous line) and to the CTA goal (black dashed line) and requirement (red continuous line) at 125~MHz/pixel of night sky background. Right: The pulse maximum amplitude distribution for different mean detected charge (6, 90 and 265 p.e.) shows the evolution of the SiPM response to various light level. The conversion factor for these measurements was about 10~LSB/p.e. If at low photon count the single photons can be clearly observed, at large photon count, the spread from avalanche to avalanche into the micro-cells blurs the detector response.}
	\label{fig:ChargeRes}	
\end{figure}

An intensive campaign of preliminary measurements on single components (light concentrators, sensors, electronics boards, etc.) and on fully assembled modules has been devoted to assess the validity of the design, to test the components and to derive a first evaluation of the camera performance. Different figures of merit (e.g. transmittivity of cones, dark count rate and cross talk of sensors, full functional tests of the electronics and optical tests of the modules (see Refs. \cite{CONES,CameraPaper, MPPC, ElectronicsPaper}) have been measured systematically in dedicated setups. An important figure of merit of the full system (optical system plus full readout chain) is the charge resolution, for which CTA dictates strict requirements on signals from 1 to 2000 photo-electrons per pixel. This quantity has been measured on fully assembled modules connected with a prototype DigiCam system in order to reproduce the same conditions of operation of the final camera prototype. Details on the instrumentation that has been used and on the data analysis that has been carried out can be found in \cite{CameraPaper}. The result, shown in figure \ref{fig:ChargeRes}, demonstrates that the SST-1M camera performs better than the foreseen goal at different night sky background conditions. Even with a night sky background of 660 MHz per pixel, which corresponds to nights with half moon, full compatibility with the requirements is achieved. The feature appearing in the charge resolution curve at around 1000 photo-electrons is due to the transition of the system to a regime where the preamplifier is in saturation (see \cite{CameraPaper}).

\section{Auxiliary systems}
\label{sec:Aux}
The main auxiliary systems described in \cite{SPIE_Enrico} are:
\begin{itemize} 
	\item the pointing system that ensure a pointing accuracy of 7~arcsec;
	\item the camera server that operates the bridge between the camera and the array control system including the data acquisition chain;
	\item the safety system that which ensure the safe and proper operation of the telescope
	\item the control cabinet hosting the services next to the telescope
\end{itemize}
\section{Conclusion}
\label{sec:concl}
The SST-1M prototype is being developed and has reached the early phase of commissioning. has been conceived for providing the required sensitivity for the high energy gamma ray program of the CTA project. A telescope prototype is being developed which fully complies with the CTA requirements in terms of technical and physical performance. Its design has been carried out in view of a possible production of up to 20 units, and resulted in a light and compact structure that can be reproduced using standard industrial procedures. Preliminary tests have fully validated the performance of the SST-1M. First sky observations are foreseen before the end of 2016, first at the Observatory of the University of Geneva followed by its installation onto the telescope prototype structue in Krak\'ow.

\section{Acknowledgments}
We gratefully acknowledge support from the University of Geneva, the Swiss National Foundation, the Ernest Boninchi Foundation and the agencies and organizations listed under Funding Agencies at this website: http://www.cta-observatory.org/
In particular we are grateful for support from the NCN grant DEC-2011/01/M/ST9/01891 and the MNiSW grant 498/1/FNiTP/FNiTP/2010 in Poland.

\nocite{*}
\bibliographystyle{aipnum-cp}%
\bibliography{Gamm16_Heller}%

\end{document}